\documentclass[12pt]{article}
\usepackage[T1]{fontenc}
\usepackage{a4wide}
\usepackage{graphics}
\usepackage{amssymb}
\usepackage{eepic}
\usepackage{psfrag}
\usepackage{epsf}

\newcommand{\subsubsubsection}[1]{{\bf #1}}

\makeatletter
\renewcommand\theequation{\hbox{\normalsize\arabic{section}.\arabic{equation}}}
\@addtoreset{equation}{section}
\makeatother
\newcommand\unt{unitary}
\newcommand\real{real}
\newcommand\nreal{nonreal}

\def\aoo{{\ensuremath{a_1^{(1)}}}}
\def\att{{\ensuremath{a_2^{(2)}}}}

\def\cM{{\ensuremath{\mathcal M}}}
\def\qp{{\ensuremath{\arg(q)/\pi}}}
\def\QP{{\ensuremath{\frac{\arg(q)}\pi}}}

\begin{document}

\def\today{February 8, 2002}

\title{RSOS revisited}

\author{G.~Tak\'acs\thanks{
E-mail: takacs@ludens.elte.hu
} ~and G.M.T.~Watts\thanks{
E-mail: gmtw@mth.kcl.ac.uk}\\
\\
*
Institute for Theoretical Physics, E\"otv\"os University,\\
P\'azm\'any P\'eter s\'et\'any 1/A,
Budapest H-1117
\\
\\
$\dagger$
Department of Mathematics, King's College London \\
Strand, London WC2R 2LS, United Kingdom
}

\maketitle

\begin{abstract}
We investigate the issues of unitarity and reality of the spectrum for
the imaginary coupled affine Toda field theories based on \aoo\ and
\att\ and the perturbed minimal models that arise from their various
RSOS restrictions. We show that while all theories based on \aoo\ have
real spectra in finite volume, the spectra of \att\ models is in
general complex, with some exceptions. We also correct the $S$
matrices conjectured earlier for the $\phi_{15}$ perturbations of
minimal models and give evidence for a conjecture that the RSOS
spectra can be obtained as suitable projections of the folded ATFTs in
finite volume.
\end{abstract}
\vspace{1cm}
hepth/0203073
\\
Report-no: KCL-MTH-02-04, ITP-Budapest-579

\newpage

\section{Introduction}

Unitarity plays an important role in quantum field theory (QFT). In
many application of QFT, the framework of quantum theory requires a
positive definite conserved probability, which is guaranteed by the
Hermiticity of the Hamiltonian and the unitarity of the $S$
matrix. Hermiticity also guarantees that the spectrum of the
Hamiltonian is real, an important condition if the Hamiltonian is to
be interpreted as the energy of the physical system. 

However, there are applications of the QFT formalism when this is not
a necessary physical requirement: non-unitary QFTs (even those with
complex spectra) appear to play an increasing role in the
investigation of statistical mechanical systems (e.g. disordered
systems \cite{disorder}).

In this paper we intend to study the simplest cases: the finite-size
spectra of models related to imaginary coupled Toda theories \aoo\
(sine-Gordon) and \att\ (ZMS), with periodic boundary conditions. Our
main aim is to establish conditions under which these theories have
real spectra, continuing our earlier work started in
\cite{toda_unitarity}. Reality of the spectrum could be interesting
for several reasons: (i) it affects the large distance asymptotics of
correlation functions (in case of a complex spectrum, these
asymptotics can show oscillating behaviour), (ii) it could allow a
redefinition of the theory that renders it unitary 
(although, as we discuss, physical requirements may prevent that,
and it is not at all clear whether this can be performed while
maintaining the interpretation of the given model as a local
QFT). Finite size spectra are also interesting in their own right:
periodic boundary conditions can also be thought of as realization of
finite temperature.

We start by recalling the general issues involved, and the
relationships between the three distinct concepts of probability
conservation, unitarity and the reality of the spectrum.

For any imaginary coupled affine Toda field theory (ATFT), there are
three different classes of models which can be considered, namely the
\emph{original (unfolded, unrestricted) 
}, the \emph{folded} and the
\emph{restricted (RSOS)} models.  We define these in section
\ref{sec:models} and discuss the reality of the spectrum in these
three classes for the theories related to \aoo\ (sine-Gordon) and
\att\ (ZMS) in turn.

In the case of \aoo, the finite-size spectrum of the RSOS models turns
out to be a \emph{subset} of the spectrum of a suitable folded model
-- i.e. the state space of the RSOS model can be thought of as a
projection of the space of the folded model. Using this relation, we
can conclude that the spectra of all \aoo\ related models are real.
In this paper we also give evidence that a similar relationship exists
between the RSOS and folded models based on \att. However, this is not
sufficient to establish the reality of these models. We then turn to
transfer matrix arguments and numerical analysis to investigate the
spectrum.

In all cases in which our studies find no violation of reality, there
are several possibilities remaining:

\begin{enumerate}
\item{} Apart from cases in which the $S$-matrix can be shown to be
Hermitian analytic (e.g. perturbations of unitary minimal models), we
only have numerical (and sometimes matrix perturbation theory)
results. Since these numerical tests require `scanning' over the
full range of rapidity of each particle in a multi-particle state,
there is always a chance that such methods miss some region of
rapidities in which reality is violated. In addition, numerical
diagonalisation always introduces some numerical errors, and so some
threshold condition must be defined to separate real cases from
nonreal cases. It remains a question whether reality violations below
the threshold are genuine or not.
\item{} We could only examine transfer matrices containing up to five
particles, as the numerical calculations become progressively more and
more difficult as we increase the number of particles
involved. Therefore, it is possible that higher particle transfer
matrices would introduce further constraints on the reality of the
spectrum.
\item{} In the case of \att, our methods only allow us to determine
the large volume asymptotics of the spectrum which have a power-like
decay as a function of the volume. It is possible that further
contributions (decaying exponentially with the volume, e.g. those
related to vacuum polarisation) would spoil reality.
\item{} For \att, the bootstrap is closed only in certain regimes of
the parameter space and we do not know the $S$-matrices of all
possible particles in the spectrum. It is possible that reality of the
spectrum is only violated in sectors which contain such `extra'
particles.
\end{enumerate}
As a result, we can only have definite results in the cases when the
spectrum is complex: if we find that some multi-particle transfer
matrix has non-phase eigenvalues then we can conclude that the
Hamiltonian must have complex eigenvalues. Whenever our examination
finds that the spectrum is real we cannot the possibility exclude that
further study would find a complex spectrum. 

However, the theories for which we cannot find any violation of
reality show some very striking patterns, and therefore we believe it
is possible that several or possibly all theories falling into these
patterns have real spectra. Clearly, further understanding is
necessary: the most promising approach would be the development of
some exact method to describe the finite size spectra (e.g. an
extension of the \att\ NLIE \cite{zms_nlie} to describe excited
states) and a deeper understanding of the relation between the folded
and the RSOS models in finite volume.

As a side result, we also correct some minor
mistakes in our previous paper \cite{toda_unitarity} and the
$\phi_{15}$ $S$-matrices conjectured earlier by one of us in
\cite{rsos15}.  We summarise our results in section
\ref{sec:conclusions}.

\section{Unitarity and related concepts in QFT}

In this section we discuss the relation between the concepts of
probability conservation, unitarity and reality of the spectrum in
quantum field theory which are often confused.
We also recall the definitions of Hermitian analyticity and R--matrix
unitarity and their relationship with unitarity of the $S$-matrix.

\subsection{Unitarity, reality and probabilistic interpretation}

For all the theories we consider there exists a non-degenerate
sesquilinear form (which we call an inner product for short, even if
it is not positive definite) on their space of states which is
conserved under the time evolution described by the Hamiltonian
i.e. the time evolution operator and the $S$-matrix preserves this
inner product (and the Hamiltonian is Hermitian with respect to the
conjugation defined by it). For perturbed conformal field theories
such an inner product is inherited from the standard inner product
used in CFT, even away from the critical point.

In the cases when this inner product is positive definite (and defines
a Hilbert space structure on the space of states) this implies the
usual Hermiticity/unitarity, with the consequence that the $S$-matrix
has phase eigenvalues and the energy spectrum is real. This inner
product makes possible the usual probabilistic interpretation in
quantum theory. These theories are called \emph{\unt} QFTs.

On the other hand, in many systems the inner product is indefinite. It
is still possible for the spectrum to be real and the $S$-matrix
eigenvalues to be phases in which case we call the theory a
\emph{\real} QFTs. However, generally when the inner product is
indefinite, then the spectrum is complex and the eigenvalues of the
$S$-matrix are not phases, in which case we call the QFT
\emph{\nreal}\footnote{%
There doesn't seem to be any agreed convention for the naming of what
we call \real\ and \nreal\ theories. We simply use these names in this
paper for convenience.}.

One can see that the three notions of unitarity, reality and
probability conservation are in fact different. If we want the theory
to have a \emph{positive definite conserved probability}, then
unitarity is necessary, which is a stronger notion than reality.
However, often this is not required; moreover, there exist
exotic probability theories which allow for negative or complex
probabilities (which in that case of course loses the interpretation
in terms of frequency of events).

An important point is whether the $S$-matrix has any relevance in the
case when the theory is non-unitary. It is easier to accept this when
the theory is still real, as in the case of the scaling Lee-Yang model
(i.e. Virasoro minimal model $(2,5)$ perturbed by its single
nontrivial relevant operator $\phi_{13}$)
\footnote{%
For an early discussion of this issue,
cf. \cite{cardy_mussardo}.  } 
or for indeed of any $\phi_{13}$
perturbations of Virasoro minimal models.  However, there are examples
of \nreal\ theories (e.g. the Virasoro minimal model $(3,14)$
perturbed by operator $\phi_{15}$ \cite{12_15}) for which the finite
size spectrum extracted using thermodynamic Bethe Ansatz (TBA) for
the vacuum energy and the Bethe-Yang equations for multi-particle
states matches perfectly the spectrum of the Hamiltonian obtained
numerically using truncated conformal space approach (TCSA), even for
the complex part of the spectrum. 

To summarise, we believe that for a large class of theories (unitary,
real and non-real), and in particular for the RSOS models considered in
this paper, the spectrum is determined by the S-matrix even when 
the spectrum is complex.

\subsection{Unitarity and Hermitian analyticity}

In analytic $S$-matrix theory the property of unitarity
is closely linked with that of Hermitian analyticity
\cite{olive,miramontes}. Without entering into details, we intend to recall
these concepts and their relations here as they are going to play an
important role later.

Unitarity is simply the statement that using a Hermitian conjugation
$\dagger$ with respect to a positive definite inner product, the $S$
operator that maps out-states into the in-states has the property
\begin{equation}
SS^{\dagger}=S^{\dagger}S={\mathbb I}\,. \label{unitarity}
\end{equation}
Given a theory that is non-unitary but real, since the $S$ operator has
phase eigenvalues it is obviously possible to define a new positive
definite inner product with respect to which $S$ is unitary. However,
this inner product may be inconsistent with the rules of analytic $S$
matrix theory and/or may render the Hamiltonian non-Hermitian (while
the Hamiltonian was Hermitian with respect to the original, indefinite
inner product). The scaling Lee-Yang model is an example of this
situation. Its spectrum contains a single scalar particle which
appears as a bound state in the two-particle scattering. $S$-matrix
theory then relates the norm of one-particle states to two-particle
states through the residue of the corresponding pole in the
two-particle $S$-matrix. The natural inner product in the perturbed
CFT formalism (defined through the Hermiticity of Virasoro generators)
is indefinite, with $n$-particle states having the signature $(-1)^n$
(or $-(-1)^n$, depending on choice, in which case the sign of the
residue is consistent with $S$-matrix theory. However, if one attempts
to use a positive definite inner product, then the residue of the
two-particle $S$-matrix at the bound state pole has the ``wrong
sign'', which is how it is often stated in the literature. As a
result, the natural inner product of the scaling Lee-Yang model is
indefinite, and it is an example of a \emph{real} but
\emph{non-unitary} theory. Therefore the existence of a positive
definite redefinition of the inner product does not mean the theory
can be made unitary because it may conflict with some other physically
motivated requirements. For a further discussion of this issue see
\cite{cardy_mussardo}.
The property that  
the residues of the bound state poles have the ``right sign'' is
sometimes called ``one-particle unitarity'' \cite{one_p_unit}.

In integrable theories the whole $S$ operator is encoded in the set of
two-particle $S$-matrices $S_{AB}(\theta)$, where $A$ and $B$ denote
the particles (or multiplets if there are internal quantum numbers)
and $\theta$ is their relative rapidity. It is a simple matter to
prove that unitarity of all possible two-particle $S$-matrices
\begin{equation}
\sum_{k,l}S_{AB}(\theta)_{ij}^{kl}\left(S_{AB}(\theta)_{mn}^{kl}\right)^{*}=
\delta_{im}\delta_{jn}\,,
\label{TU}\end{equation}
(where we explicitly wrote the multiplet indices) implies the
unitarity of multi-particle transfer matrices (defined in Appendix
\ref{transfermat}), and also means that the spectrum is real. In
writing equation (\ref{TU}) we assumed that we had chosen an orthonormal
basis in the internal multiplet space.

If, in addition, the inner product on the space of states is positive
definite then the theory in question is a unitary QFT. We shall
abbreviate the property (\ref{TU}) as TU (two-particle unitarity).  

Hermitian analyticity (HA) tells us something about the behaviour of the
$S$-matrix elements under complex conjugation. It states that 
\begin{equation}
\left(S_{AB}(\theta)_{ij}^{kl}\right)^{*}=S_{BA}(-\theta^*)_{lk}^{ji}\ .
\label{HA}\end{equation}

On the other hand, the $S$-matrices we investigate here are derived
from quantum group $R$-matrices. $R$-matrices also satisfy a relation
known as ``unitarity'' in quantum group theory which we call here
$R$-matrix unitarity (RU) and takes the form

\begin{equation}
\sum_{k,l}S_{AB}(\theta)_{ij}^{kl}S_{BA}(-\theta)_{lk}^{nm}=
\delta_{im}\delta_{jn}\,.
\label{RU}\end{equation}

We see that RU and HA together imply TU and thus reality of the
spectrum (note that equation (\ref{TU}) is meant only for physical
i.e. real values of $\theta$). Therefore for the models we consider
Hermitian analyticity implies a real spectrum since RU is automatically
satisfied due to the quantum group symmetry.

These notions can be appropriately generalised to RSOS theories, where
the multi-particle polarisation spaces are not simply tensor products
of one-particle ones. Rather, the multiplet structure is labelled
by so-called RSOS sequences, which denote `the vacua' between which the
particles mediate and are constrained according to so-called
`adjacency rules'. The two-particle $S$-matrix therefore carries
four vacuum indices $a,b,c,d$ and it is specified in the following way
(we omit particle species labels for simplicity):
\begin{equation}
\setlength{\unitlength}{2500sp}%
\begingroup\makeatletter\ifx\SetFigFont\undefined%
\gdef\SetFigFont#1#2#3#4#5{%
  \reset@font\fontsize{#1}{#2pt}%
  \fontfamily{#3}\fontseries{#4}\fontshape{#5}%
  \selectfont}%
\fi\endgroup%
\begin{picture}(1440,1455)(451,-541)
\thinlines
{\put(901,-601){\line( -1,0){100}}
}%
{\put(901,-601){\line( 0,-1){100}}
}%
{\put(1621,-601){\line( 1,0){100}}
}%
{\put(1621,-601){\line( 0,-1){100}}
}%
{\put(901,119){\line( 1,0){100}}
}%
{\put(901,119){\line( 0,-1){100}}
}%
{\put(1621,119){\line( -1,0){100}}
}%
{\put(1621,119){\line( 0,-1){100}}
}%
{\put(721,299){\line( 1,-1){1080}}
}%
{\put(721,-781){\line( 1, 1){1080}}
}%
\put(721,-331){\makebox(0,0)[lb]{\smash{\SetFigFont{12}{14.4}{\rmdefault}{\mddefault}{\itdefault}{a}%
}}}
\put(1216,-781){\makebox(0,0)[lb]{\smash{\SetFigFont{12}{14.4}{\rmdefault}{\mddefault}{\itdefault}{b}%
}}}
\put(1666,-331){\makebox(0,0)[lb]{\smash{\SetFigFont{12}{14.4}{\rmdefault}{\mddefault}{\itdefault}{c}%
}}}
\put(1171,119){\makebox(0,0)[lb]{\smash{\SetFigFont{12}{14.4}{\rmdefault}{\mddefault}{\itdefault}{d}%
}}}
\put(1891,-1141){\makebox(0,0)[lb]{\smash{\SetFigFont{12}{14.4}{\rmdefault}{\mddefault}{\itdefault}{$\phi$}%
}}}
\put(451,-1141){\makebox(0,0)[lb]{\smash{\SetFigFont{12}{14.4}{\rmdefault}{\mddefault}{\itdefault}{$\theta$}%
}}}
\end{picture}
\;\;\longleftrightarrow\;\;
  S^{cd}_{ab}(\theta-\phi)
\;.
\end{equation}
\vspace{0.1cm}

For RSOS theories TU, HA and RU take the form:
\begin{eqnarray}
&&\mathrm{TU}:\quad \sum_{d}S_{ab}^{cd}(\theta)S_{ae}^{cd}(\theta)^*=\delta_{be}\label{TU_RSOS}\\
&&\mathrm{HA}:\quad
S_{ae}^{cd}(\theta)^*=S_{ad}^{ce}(-\theta^*)\label{HA_RSOS}\\
\nonumber \\
&&\mathrm{RU}:\quad \sum_{d}S_{ab}^{cd}(\theta)S_{ad}^{ce}(-\theta)=\delta_{be}\quad \label{RU_RSOS}
\end{eqnarray}

\section{The models}
\label{sec:models}

The models we study can be classified in the following way:
\begin{enumerate}

\item{} 
\emph{The original 
(unfolded, unrestricted) 
models}. 
These have an infinite set of degenerate `vacua' and their spectra are
built from a fundamental soliton doublet (\aoo) or triplet (\att) by
closing the $S$-matrix bootstrap. 

\item{} 
\emph{Folded models}. 
Using the periodicity of the field theoretic potential, one can choose
to identify the ground states  
after $k$ periods (see \cite{ksg} for the sine-Gordon case), i.e. in a
$k$-folded model one has $k$ ground states, between which the solitons
mediate. The spectrum and the scattering theory for this case can be
straightforwardly written down using the well-known $S$-matrices of
the original model.

\item{} 
\emph{Restricted (RSOS) models}.  
At certain `rational' values of the coupling it is possible to define
a space of `RSOS states' as a quotient of the full space of states on
which the action of the $S$ operator is well defined. These states can
be labelled by sequences of `vacua' and the $S$-matrix factories into
$2$-particle `RSOS type' $S$-matrices. These RSOS $S$-matrices
describe the scattering theory of the $\phi_{13}$ or
$\phi_{12}$/$\phi_{21}$/$\phi_{15}$ perturbations of Virasoro minimal
models, respectively for \aoo\ and \att. These restrictions were
discussed in the following papers: the $\phi_{13}$ case in
\cite{reshetikhin_smirnov}, $\phi_{12}$/$\phi_{21}$ in \cite{smirnov}
and $\phi_{15}$ in \cite{rsos15}. The above construction of the RSOS
states relies on the action of the quantum group and has only been
defined for the theory on the full line i.e. in infinite spatial
volume. However, given the two-particle $S$-matrix derived in this way
one can easily define the theory in finite volume.
\end{enumerate}

All of the models enumerated above are \emph{integrable} and therefore
there are numerous results for their finite volume spectra. We shall
mainly use 
results obtained using two methods. 

Firstly, the so-called
\emph{nonlinear integral equation (NLIE)} approach, which gives exact
results for the spectrum and was pioneered by Kl\"umper et
al. \cite{KBP} on the one hand and Destri and de Vega \cite{DdV-92-95}
on the other. The original developments concerned mainly the spectra
of the \aoo\ case (for the excited states see also
\cite{FMQR,DdV-97,frt}). 

Secondly, we use the related method of the Bethe-Yang equations
(cf. \cite{klassen_melzer1} and references therein), which describe
the large volume asymptotics of the finite size spectrum. If the NLIE
for a given theory is known, then the asymptotics can be derived
independently and for \aoo\ it was checked that they agree with the
results from the Bethe-Yang equations \cite{DdV-97,frt}. In general
when the NLIE is not known this is the only known analytic way to
obtain information about the finite size spectrum (in some cases, TBA
equations are known for the excited states \cite{excited_tba}, but
these cases are covered by the NLIE as well).

In finite volume, since the field theory potential is periodic, it is
possible to introduce twisted sectors
 of the original model 
\cite{polymer}. These are labelled by a twist parameter $\alpha$,
defined $\bmod\,2\pi$. The twisted sectors have different finite size
spectra which are given by a modification of the original NLIE in
which $\alpha$ appears explicitly \cite{polymer}.

The finite size spectrum of the $k$-folded model is the union of the
spectra of twisted sectors where the twist runs
through the values \cite{ksg}

\begin{equation}
\alpha = \frac{2\pi m}{k}\quad ,\quad  m = 0,...,k-1
\end{equation}

When we turn to specific models we shall discuss a relation between
the finite size spectra of folded and RSOS models,
namely that the finite-size spectrum of an RSOS model is exactly a
subset of the spectrum of an appropriate folded model.
Such a relation first emerged for the case of \aoo\ \cite{frt_alpha},
and we shall give evidence that a similar relationship holds for
\att.

To be  precise, for \aoo\ it is known that both the exact spectra
(described by the NLIE) and, as a consequence, their asymptotics
(described by the Bethe-Yang eqns) of folded and RSOS models are
related. For \att\ it is known that the exact vacuum energies in
finite volume are related (this was shown using the NLIE framework in
\cite{zms_nlie}), but at present the NLIE for excited states is not
known. We shall present evidence that the asymptotics of the excited
state energies given by the Bethe-Yang eqns are similarly related. We
believe similar relations will hold for the exact spectra of every
ATFT and the corresponding RSOS models. 

There exists some evidence in support of this claim.
Al.B.~Zamolodchikov has shown that within the perturbed conformal
field theory framework, the perturbative expansion for the vacuum
energy of the sine-Gordon model with a suitable value of the twist
parameter could be reinterpreted as a perturbative expansion in the
RSOS model \cite{polymer}, to all orders in perturbation theory.
There is also a quantum group argument for the agreement of the vacuum
energies \cite{PZJ}, which relies on a formulation of the partition
function in terms of the system in infinite volume. In addition, using
the \aoo\ NLIE it is possible to calculate (a) the exact ultraviolet
conformal weights and (b) the energies of the ground state and excited
states in the twisted sectors of sine-Gordon model to very high
numerical accuracy. Comparison of these data to (a) the known CFT data
and (b) numerical finite volume spectra extracted using TCSA,
respectively, shows an excellent agreement \cite{frt_alpha}.

It should be possible to find a proof using quantum group arguments
directly for the theory in finite volume, by finding an appropriate
projection from the folded model to the RSOS space which commutes with
the transfer matrices, but we are unaware of such an argument. The
knowledge of this projection would also be useful because it could be
used to select systematically the RSOS states in the NLIE approach.

We now proceed to give our conventions for the models based on \aoo\ 
and \att\ and give the RSOS---folded relations that we shall check
in Section \ref{sec:results}.

\subsection{Conventions for \aoo}

The \aoo\ theory is simply the sine-Gordon model, for which we take
the action to be

\begin{equation}
{\mathcal A}_{\mathrm sG}=\int{\mathrm d}^2x\left(
\frac{1}{2}\partial_\nu\Phi\partial^\nu\Phi+
\frac{m_0^2}{\beta^2}\cos\beta\Phi
\right)\, .
\end{equation}

As is well known, introducing the parameter $q=\exp(8\pi i/\beta^2)$,
one can show that the model is invariant with respect to $U_q(\aoo)$.
The spectrum consists of a doublet of solitons and a collection of
scalar particles (breathers).
As described by \cite{reshetikhin_smirnov}, the RSOS theory ${\mathcal
M}_{p,p'}+\phi_{13}$ is obtained as a restriction at
$\beta=\sqrt{8\pi p/p'}$, that is at $q=\exp(\pi i p'/p)$.
Putting together the results of \cite{frt} and \cite{ksg}, it is
straightforward to see that the finite volume spectrum of this RSOS model is a
subset of that of the $2p$-folded model.

\subsection{Conventions for \att}

We take the action of the $a_2^{(2)}$ model to be

\begin{equation}
{\cal A}_{\mathrm ZMS}=\int{\mathrm d}^2x
\left[
\frac{1}{2}\partial_\nu\Phi\partial^\nu\Phi+
\frac{m_0^2}{\gamma}\Big(
\exp\left(2i\sqrt{\gamma}\Phi\right)+2\exp\left(-i\sqrt{\gamma}\Phi\right)
\Big)
\right]\,.
\end{equation}

Introducing the parameter
\begin{equation}
q=\exp\left(\frac{i\pi^2}{\gamma}\right)
\label{eqn:qdef}\end{equation}
one can show that the model has a symmetry under the affine quantum
group ${\mathcal U}_q(a_2^{(2)})$ and as a result its
$S$-matrix can be explicitly constructed \cite{smirnov}.

The spectrum of the original model consists of a fundamental soliton
transforming as a triplet under $U_q(\aoo)$, having an $S$-matrix of
the form
\begin{equation}
  S^0_{00}(x,\gamma) \, R(x,q)
\;,\quad
  \hbox{where $x=\exp( 2 \pi \theta / \xi)$ and $
 \xi = \frac{2\pi}3\frac{\gamma}{(2\pi - \gamma)}$}\,,
\end{equation}
where $S^0_{00}$ is a scalar  function which is always a pure phase
for real $x$ and $\gamma$, and $R(x,q)$ is the $R$-matrix of
$U_q(\att)$ in the triplet representation. 

Depending on $\gamma$, there may be other particles in the spectrum.
For $\xi < \pi$ there are breathers $B^n_0\,,\,n=1,\dots,[\pi/\xi]$
formed as bound states of $K_0$ transforming in the singlet
representation, and for $\xi < 2 \pi/3$ there are higher kinks
$K_i\,,\,n=1,\dots,[2\pi/(3\xi)]$, also formed as bound states of
$K_0$ and transforming in the triplet representation, and their
associated breathers $B^n_i$.  

In addition, it is known that there are values of $\gamma$ for which
the bootstrap does not close on this particle content, and for which
there are more particles in the spectrum.  Since we are only able to
investigate the transfer matrices for the kinks $K_i$ and breathers
$B^n_i$, it is only possible to prove that any particular model has a
complex spectrum. If we find that the spectrum for these particles is
real, that does not imply the reality of the full spectrum simply
because the $S$--matrix bootstrap could still generate $S$--matrices
with non-phase eigenvalues.

The $S$--matrices of the higher kinks have the form
\begin{equation}
  S_{K_m,K_n}(\theta)
= S^0_{m,n}(x,\gamma)\, R(\,(-1)^{m+n}x\,,q)
\,
\end{equation}
where $S^0_{mn}$ is a scalar factor.

Since the scalar factors $S^0_{m,n}$ are phases and there are only two
different matrix structures for the two-particle $S$-matrices
$S_{K_m,K_n}$ i.e $R(x\,,q)$ if $m+n$ is even and $R(-x\,,q)$ if $m+n$
is odd, for the purposes of finding out if the eigenvalues of the
transfer matrices (defined in appendix \ref{transfermat}) it is hence
to sufficient to consider
\begin{itemize}
 \item{For $\pi/\gamma < 1$:} transfer matrices containing only $K_0$  
 \item{For $\pi/\gamma > 1$:} transfer matrices containing only $K_0$
 and/or $K_1$ 
\end{itemize}

\subsubsection{RSOS models related to \att}

The \att\ model has two inequivalent RSOS restrictions
\cite{smirnov,rsos15}. 
{}From the conjectured ground state NLIE of the $a_2^{(2)}$-related
models \cite{zms_nlie} one can determine the minimal value of the
folding number for which the RSOS ground state is in the spectrum of
the folded model. Together with the results in \cite{smirnov,rsos15},
it leads to the following {conjectures} 
\begin{enumerate}
\item{}\emph{$\cM_{p,p'}+\phi_{12}$ is a projection of the
$p$-folded ZMS model at $\gamma=\pi p/p'$}. 

The RSOS $S$-matrices are
given in the appendix \ref{rsos12} where we have specified all the
formulae including the $6j$ symbols since the ones we found in the
literature all had typos.\footnote{%
We are grateful to Giuseppe Mussardo (SISSA) for
providing us with a correct set of formulae.}

As for the unrestricted case, the matrix part of the $S$-matrices
depend only on $q$ so they depend on $p'$ only modulo $2p$. 
In fact, we only need study the values $1\leq p'<p$ since the substitution
$p'\,\rightarrow\,-p'$ changes the $S$-matrix elements to their
complex conjugate (if we choose the square root branches in the $6j$ symbols
appropriately) and thus simply complex conjugates the transfer matrices and 
their eigenvalues too as well.

(n.b. the models $\cM_{p,p'}+\phi_{21}$ are contained in
this class through the identification ${\mathcal
M}_{p,p'}+\phi_{21} \equiv \cM_{p',p}+\phi_{12}$.)

\item{}\emph{$\cM_{p,p'}+\phi_{15}$ is a projection of the
$2p$-folded ZMS model at $\gamma=4\pi p/p'$}. 

The RSOS restriction
leading to the $S$-matrices was performed in \cite{rsos15}, however,
certain amplitudes had wrong prefactors. The reason was that in the
scattering amplitudes of two (oppositely) charged solitons into two
neutral ones there is a Clebsh-Gordan factor which was not taken
properly into account and therefore the amplitudes of \cite{rsos15} do
not satisfy the Yang-Baxter equation and $R$-matrix unitarity. The
corrected amplitudes are listed in Appendix \ref{rsos15}.
\end{enumerate}
These perturbations lead to renormalisable field theories if the
conformal weight of the perturbing field is less than or equal to two,
which is equivalent to the condition $\pi/\gamma \geq 1/2$ for both
cases, which is also the condition for the unrestricted model to be
well-defined.

\section{Results}\label{sec:results}

The leading approximation (e.g.\ in the NLIE formalism) to finite size
effects in large volume is given by the Bethe-Yang equations which are
summarised in Appendix \ref{transfermat}. In particular it can be seen
that the spectra of the models can only be real if the transfer
matrix eigenvalues (denoted in the Appendix by $\lambda ^{(s)}\left(
\vartheta |\vartheta_{1},\ldots ,\vartheta_{N}\right)$) are all
phases for real values of the particle rapidities $\vartheta_i$.

We check for which values of the coupling the various transfer
matrices have phase eigenvalues. There are some simplifications we can
make.

\begin{enumerate}
\item{} In all our calculation we omit the scalar prefactors of the
$S$-matrices since these are irrelevant for determining whether the
eigenvalues are pure phases or not. We call the transfer matrices
obtained from the $S$-matrices without the scalar factor
`reduced' transfer matrices.
\item{} The $R$-matrices of all models we consider have some discrete
symmetries (which we describe here for real values of the rapidity).
\begin{enumerate}
\item{} $R(x,q^*)=R(x,q)^*$, which means that we only need to consider
$0\leq\arg(q)\leq\pi$.
\item{} $R(x,-1/q)=UR(x,q)U^{-1}$ where $U$ is a diagonal matrix whose
nonzero entries are $\pm 1$.
\end{enumerate}
\end{enumerate}

The reduced transfer matrices are still in general very complicated,
even for the original models, and therefore in our study we
diagonalised them numerically for a large set of values for the
rapidities and other parameters. In all our plots we show only the
eigenvalues of the reduced transfer matrices.

\subsection{$a_1^{(1)}$ related models}

In the case of $a_1^{(1)}$ related models, since the folded and
original models are all unitary, their finite size spectra are real
and therefore all $\phi_{13}$ perturbations have real spectra as
well, regardless of whether they are perturbations of unitary or
non-unitary minimal models. In addition, the NLIE formalism
\cite{frt_alpha} yields manifestly real spectra as well.

In other words:
\[
 \hbox{original unitary}
\Rightarrow
 \hbox{folded unitary}
\cases{
\Rightarrow
 \hbox{RSOS real} \cr
\not\Rightarrow
 \hbox{RSOS unitary}}
\]
One might think that, since the RSOS spectrum is simply a subset of
the spectrum of the unitary unrestricted model, the RSOS model would
simply inherit a positive definite inner product and also be
unitary. However, the fact that the spectrum of the RSOS and
unrestricted models are different, means that the constraints of
analytic $S$-matrix theory may enforce different inner products.  As
an example, consider the value $8\pi/\beta^2=2/5$. The particle
content of the unrestricted model is a soliton doublet $s,\bar s$ and
a single breather $B$. The RSOS model, on the contrary, is the
Lee-Yang model with a single particle $B$ (the solitons are removed
from the spectrum by the RSOS restriction). There is a first order
pole in $S_{BB}(\theta)$ at $\theta=2\pi i/3$ which must have an
explanation in terms of on-shell diagrams. In the unrestricted
sine-Gordon model, it is explained by the famous Coleman-Thun
mechanism \cite{CT} as the sum of singular contributions from 
diagrams with internal soliton lines. In the RSOS model there are no
solitons so this pole must be explained by a single diagram in which
$B$ occurs as a bound state of two $B$ particles. As we discussed
before, the sign of the residue of this pole forces the inner product
to be indefinite.

By numerical diagonalisation of the reduced transfer matrices we found
that the transfer matrix eigenvalues for the RSOS models ${\mathcal
M}_{p,p'}+\phi_{13}$ are a subset of those for the $2p$-folded
sine-Gordon model with $\beta=\sqrt{8\pi p'/p}$. We also observed
that the numerically computed transfer matrix eigenvalues are all
phases, as expected from the general argument above.

\begin{figure}[htb]
{\par\centering {\resizebox*{.7\hsize}{!}{\includegraphics{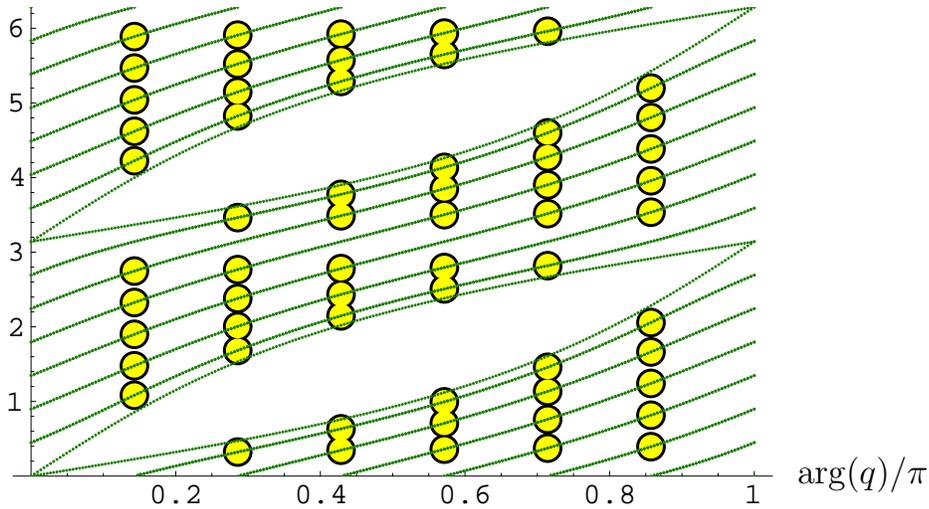}}} \par}
\caption{%
\small
Comparison of the eigenvalues of the matrix part of the
two--particle transfer matrices for the RSOS model $ (\cM_{7,m} +
\phi_{13})$ and the 14--folded \aoo\ ATFT\ The horizontal axis is
$\arg(q)/\pi$ restricted to the fundamental domain between $0$ and
$1$, and the vertical axis is the argument of the eigenvalues. The
folded models are shown in green/solid lines, and the RSOS models
(which are only defined for discrete values of $q$) are shown as
blobs. The relative rapidity of the particles was chosen $\theta=1/9$.  }
\label{fig:13}
\end{figure}

In figure \ref{fig:13}, we plot the arguments of the eigenvalues of
the reduced two--particle transfer matrices of the RSOS models $
(\cM_{7,m} + \phi_{13})$ and the 14--folded sine-Gordon model.  The RSOS
models are only defined for $m$ integer, i.e. for $\arg(q)/\pi$ taking
values $m/7$. For these values the eigenvalues of the transfer
matrices (shown as blobs) are a subset of the eigenvalues of the
matrices of the folded model shown as lines.

\psfrag{Arg}{$\arg(q)/\pi$}

The two-particle space of the 14-folded model has dimension 28, and
the spectrum consists of 12 eigenvalues each of multiplicity 2 and 4
of multiplicity 1. The RSOS model excludes the eigenvalues of
multiplicity 1, 2 of the other eigenvalues, and each of the remaining
10 eigenvalues appears with multiplicity 1 only. 

There are RSOS models $\cM_{p,p'}+\phi_{13}$ for which the
$S$-matrix is Hermitian analytic (e.g. the unitary case $p'=p+1$ or
the series $p=2$, see \cite{reshetikhin_smirnov}) which then implies
that the spectrum is real. Our arguments above, however, extend to
\emph{all} $\phi_{13}$ perturbations.

\subsection{$a_2^{(2)}$ related models}

In this section, we investigate models related to \att.
We start with the  1-folded 
model where we correct a mistake in
\cite{toda_unitarity}. We then consider multi-particle transfer matrices
and show the emergence of a pattern which seems to carry over to
the folded case and it plays an important role in the RSOS case as
well.

For all three classes of model, we must consider separately the case
$\pi/\gamma > 1$ for which there are higher kinks, and $\pi/\gamma<1$
for which there are no higher kinks and consequently fewer constraints
on the values of $\gamma$ allowed by the reality of the spectrum.

We recall that due to the symmetries of the $R$--matrix, the spectrum
of models with $-\pi<\arg(q)<0$ is obtained by conjugating the
spectrum of the model with $1/q$, and so we can restrict our attention
to models with $0\leq \arg(q)\leq \pi$.

\subsubsection{The $1$-folded models based on \att}

We shall consider first the two-particle transfer matrix, and then the
higher particle number transfer matrices in turn.  As we have
mentioned, for our investigation of the reality of the spectrum, we
can ignore the scalar prefactors in the $S$--matrices and need only
consider the `reduced' transfer matrices constructed out of the
appropriate $R$--matrices.  The two-particle `reduced transfer
matrices' are simply the $R$-matrices themselves: if the kinks being
scattered are $K_m$ and $K_n$, then the appropriate $R$--matrix is
$R((-1)^{m+n}x,q)$.

As calculated in \cite{toda_unitarity}, the eigenvalues of $R(x,q)$
are three pairs of doubly degenerate  eigenvalues,
\begin{equation}
1
\;,\;\;\;\;
\left(\frac{ 1 - q^2 \sqrt x }{q^2 -  \sqrt x } \right)
\;,\;\;\;\;
\left(\frac{ 1 + q^2 \sqrt x }{q^2  + \sqrt x } \right)
\label{eq:evals}
\end{equation}
and three eigenvalues $\lambda$ satisfying
\begin{equation}
{\renewcommand{\arraystretch}{1.4}
\begin{array}{rl}
   (x - q^4)(x + q^6)  \lambda^3
\;+\;  q^6( 2  + q^2 )( \lambda^2 + x^2 \lambda)
&\\
+\; x\,(q^2 - 1)(1 - 3 q^4 + q^8)
   ( \lambda^2 + \lambda)
&\\
-\;  q^2(  1 + 2 q^2) ( x^2 \lambda^2 + \lambda)
\;+\;   (1 - q^4 x)(1 + q^6 x)
&= 0\;.
\end{array}
}
\label{eq:cubic}
\end{equation}

As stated in \cite{toda_unitarity}, for $x$ negative, 
there are obviously eigenvalues which are not phases, except when
$q^4 = 1$. However, in \cite{toda_unitarity} we stated incorrectly 
that all the
eigenvalues were phases for $x$ non-negative and $q$ a phase,
and instead the correct result is that there are
non-phase eigenvalues for $ 0 < |\arg(q)/\pi| < 1/4$. This is due to
branch cuts that appear in the solutions of the cubic equation
(\ref{eq:cubic}) which we overlooked before.

We were not able to diagonalise any higher particle number transfer
matrices exactly. Therefore we resorted to numerical methods for the
cases of three, four, five and six particles, combined with matrix
perturbation theory in the case of three and four particles.

\vspace{2mm}
\subsubsubsection{The $1$-folded models with $\pi/\gamma > 1$}
\vspace{2mm}

If $\pi/\gamma>1$ then the model contains both $K_0$ and $K_1$.
Since the matrix part of $S_{K_0,K_1}$ is $R(-x,q)$, 
the eigenvalues are just given by (\ref{eq:evals}) and
(\ref{eq:cubic}) with $x$ replaced by $-x$, and from looking at
(\ref{eq:evals}) we see that  the two-particle transfer matrix has
non-phase eigenvalues unless $q^4=1$.
This means that, unless $q^4=1$, the theory is \emph{\nreal}.
We have not investigated the case $q^4=1$ any further.

\vspace{2mm}
\subsubsubsection{The $1$-folded models with $\pi/\gamma < 1$}
\vspace{2mm}

If $\pi/\gamma<1$ then the only particles in the spectrum are the
fundamental kinks (and the first breather for $\pi/\gamma > 2/3$ ).
In particular, the model contains no higher kinks, and the
only reduced transfer matrices to be considered are those constructed
out of $R(x,q)$ with $x$ positive.

The eigenvalues of the two-particle case have already been presented,
and show that the theory is \nreal\ for $0 < |\qp| < 1/4$
and for $3/4 < |\qp|< 1$. 

Numerical investigation of the three-particle transfer matrices show
that the theory is also \nreal\ if $1/4 < |\qp| < 1/3$
or $2/3<|\qp|<3/4$.
It turns out that the strongest violation of reality
happens for small $x$ and $y$, and one can expand the eigenvalues in a
perturbation series around $x=y=0$. Some care must be taken as the
$x=y=0$ transfer matrices have a nontrivial Jordan form, so the
Lidskii--Vishik--Lyusternik generalised perturbation theory
\cite{lidskii} must be used for the expansion. Without entering into
technicalities we only wish to mention that one obtains exactly the
same pattern as described above.

Consideration of the four-particle transfer matrix adds further
regions of non-reality, $1/3 < |\qp| < 3/8 $ and 
$5/8< |\qp|<2/3$ , and of the five-particle
transfer matrix  additional regions of 
$3/8< |\qp| < 2/5$ and $3/5< |\qp|< 5/8$.

Assuming this pattern to continue for higher particle number transfer
matrices, we are led to the conjecture that 
the eigenvalues of the $n$ particle transfer matrices  are always
phases for  
\begin{equation}
  \frac 12 - \frac1{2n}
\leq 
  \left|  \QP \right| 
\leq 
  \frac 12 + \frac{1}{2n}
\end{equation}
and at the isolated points
\begin{equation}
  \left| \QP \right| 
= \frac{1}{2} \pm \frac{1}{2m}\quad ,\quad 1\leq m \leq n\,.
\end{equation}
and for every other value of $q$ there are non-phase eigenvalues for
some value of the rapidities.

This would lead to the result that the spectrum of the $1$-folded model
is always complex except for the isolated points 
\begin{equation}
  \left| \QP \right| 
= \frac{1}{2}\left(1 \pm \frac{1}{n}\right)\label{borderlines}
\end{equation}

\subsubsection{Higher folded models based on \att}

We investigated the folded transfer matrices using numerical
diagonalisation and found that the regions of real and complex
spectrum are identical in every case to those of the $1$-folded
models, independent of the folding number.

It is obvious that the regions of complex spectrum contain those of
the $1$-folded model, since the spectrum of the folded model contains
that of the $1$-folded model, but it appears that there are no further
constraints arising from the twisted sectors. 

\subsubsection{RSOS models based on \att}

The first observation to make is that since the spectrum of the RSOS
models is a subset of the spectrum of the appropriate folded model, if
the folded model has a real spectrum then so does the RSOS model.

Next, since the RSOS spectrum is a subset of the folded spectrum, it
is possible that the RSOS spectrum is real while that of the folded is
complex. This is well known to be the case for the perturbations of
the unitary minimal models $\cM_{p,p\pm 1}+\phi_{12}$ (note that
$\phi_{15}$ is never a relevant perturbation of a unitary minimal
model), and we believe that this property is shared by many other
models. 

\begin{figure}[htb]
{\par\centering {\resizebox*{.8\hsize}{!}{\includegraphics{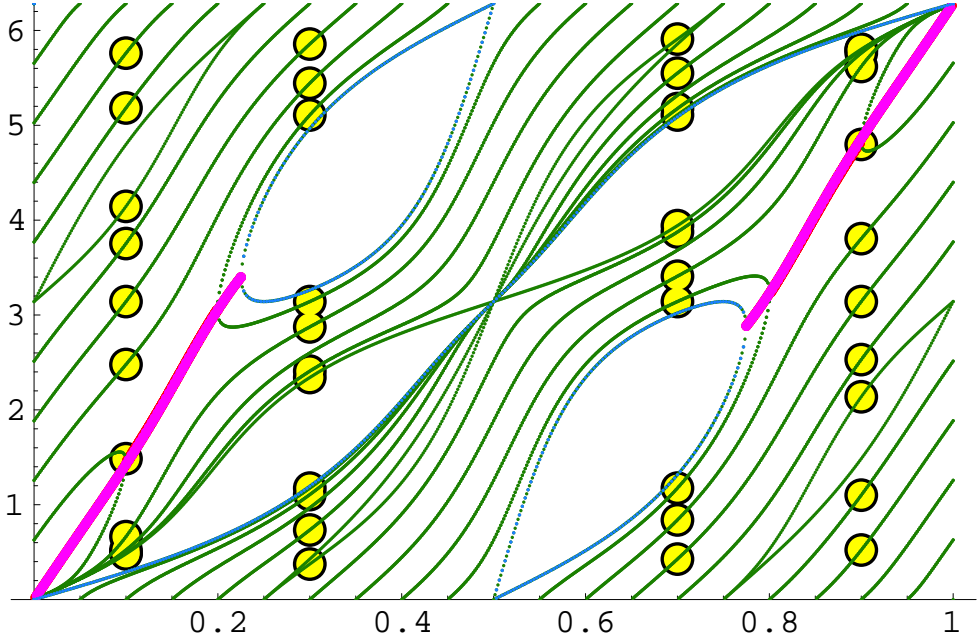}}} \par}
\caption{%
\small Comparison of the eigenvalues of the matrix part of the
two--particle transfer matrices for the RSOS model $ (\cM_{10,m} +
\phi_{12})$ and the 10--folded \att\ ATFT\ The horizontal axis is
$\arg(q)/\pi$ restricted to the fundamental domain between $0$ and
$1$, and the vertical axis is the argument of the eigenvalues. The
phase eigenvalues of the folded models are shown in green/thin solid
lines, the non-phase eigenvalues in pink/thick solid lines, and the
RSOS models (which are only defined for discrete values of $q$ and
happen in this case all to be pure phases) are shown as blobs. The
relative rapidity of the particles was chosen $\theta=5$. }
\label{fig:12}
\end{figure}

\begin{figure}[htb]
{\par\centering {\resizebox*{.8\hsize}{!}{\includegraphics{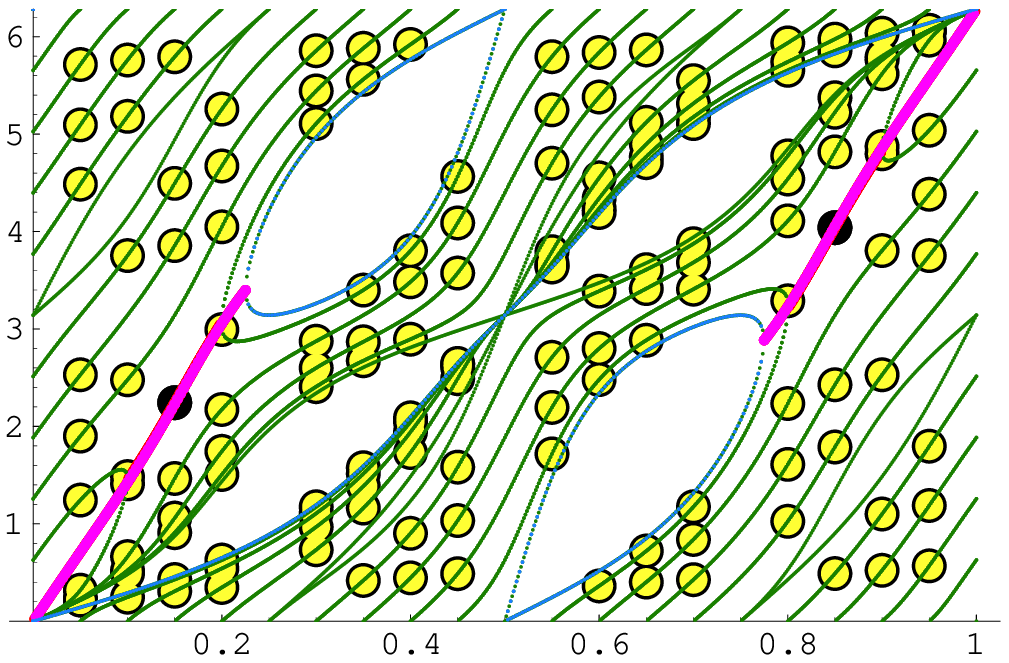}}} \par}
\caption{%
\small Comparison of the eigenvalues of the matrix part of the
two--particle transfer matrices for the RSOS model $ (\cM_{5,m} +
\phi_{15})$ and the 10--folded \att\ ATFT. The graph is labelled as
for figure \ref{fig:12},
except that there are now also non-phase eigenvalues of the RSOS model
which are shown as black blobs. The
relative rapidity of the particles was chosen $\theta=5$.}
\label{fig:15}
\end{figure}

As evidence for our assertion that the spectra of the RSOS
models are subsets of those of the folded models, in figures 
\ref{fig:12} and \ref{fig:15} we plot the eigenvalues of the reduced
two-particle transfer matrices for $K_0$--$K_0$ scattering of the
10-folded model and of the associated RSOS models
(We performed similar checks for the $K_0$--$K_1$ and
$K_0$--$K_0$--$K_0$ transfer matrices with equally convincing
results).

These figures also show that for this particular choice of rapidity
difference, there are regions of $\qp$ for which the folded
transfer matrix has non-phase eigenvalues which are included in the
regions for which we believe that the folded transfer matrices have
non-phase eigenvalues for some values of rapidity.
We also see that some of the RSOS restrictions do manage to omit these
non-phase eigenvalues, while others do not. 
We report more fully on our findings in the later sections.

\subsubsection{The $\phi_{12}$ perturbations with $\pi/\gamma > 1$}

These theories have at least one higher kink $K_1$ and we find that
the most stringent restrictions already arise from the two-particle
transfer matrix involving $K_0$ and $K_1$.

The theories with $p'=\pm 1\,\bmod\,p$ are of the `unitary type':
their $S$-matrices (and transfer matrices) are proportional to those
of the unitary models $\cM_{p,p+1}+\phi_{12}$ and
$\cM_{p,p+1}+\phi_{21}$, and since the scattering itself is manifestly
unitary the eigenvalues are all phases. Indeed these $S$-matrices do
satisfy Hermitian analyticity which together with $R$-matrix unitarity
implies the $S$-matrix unitarity equation for the two-particle
$S$-matrices.

Our numerical calculations show that the transfer matrices of the
theories $\cM_{p,p'}+\phi_{12}$ where $p'=(p\pm 1)/2\,\bmod\,p$ also
have phase eigenvalues only, although we do not know any explanation
for this fact yet.

Every other theory has non-phase eigenvalues in the $K_0$--$K_1$
transfer matrix; conversely, the models described above still have
phase eigenvalues when we consider the three-particle transfer
matrices for all combinations of $K_0$ and $K_1$.

We have not tested these models beyond the three--particle transfer
matrix. In addition, it is possible that particles arising through the
$S$--matrix bootstrap will also introduce non-phase eigenvalues in the
transfer matrices; since the bootstrap has not been completed for the
majority of these models, we cannot say anything more about them.

\subsubsection{The $\phi_{12}$ perturbations with 
$1/2 < \pi/\gamma < 1$} 

These theories are more usually known as the `$\phi_{21}$'
perturbations. They have no higher kinks in their spectra and at most
one breather, and the $S$--matrix bootstrap closes on these particles.  
The only possible violation of reality of the spectrum could be
introduced by the fundamental kink $S$--matrix, since the
kink-breather and breather-breather $S$-matrices are pure phases.

\begin{figure}[h]
{\par\centering \includegraphics{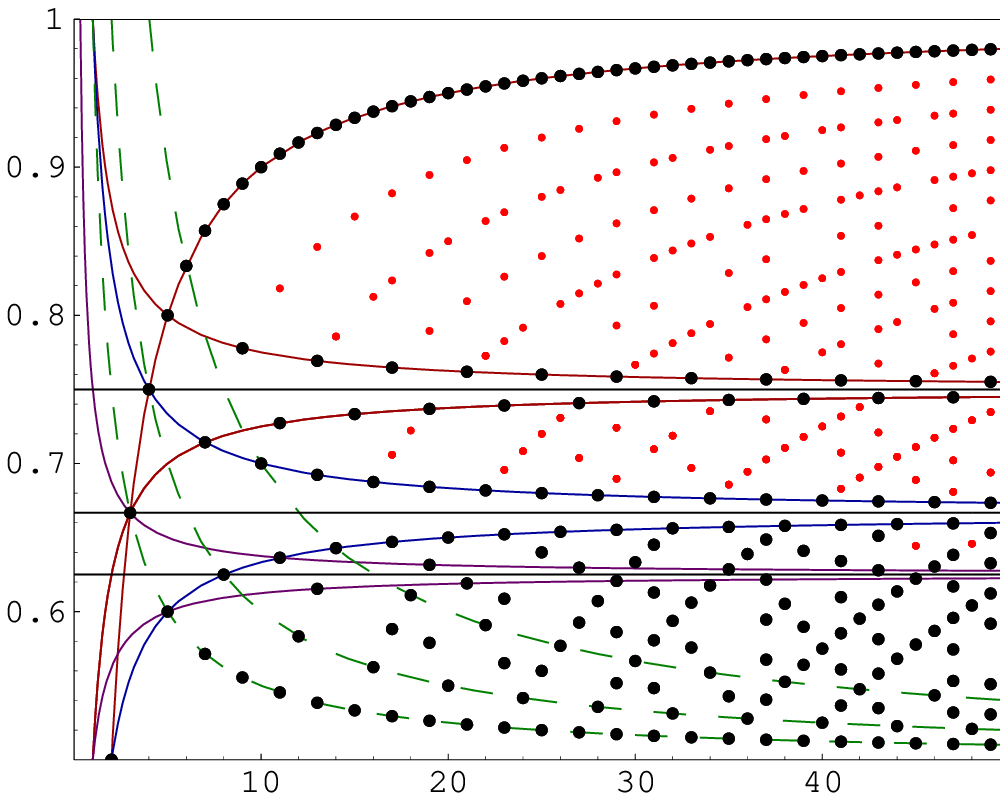} \par}

\caption{\small Numerical results for the eigenvalues of the transfer
matrices of $\cM_{rs} + \phi_{12}$.
The heavy (black) points are models for which the multi-particle
spectrum appears to be real up to 5 particles, and the light (red)
points those which are non-real, with 
$[\pi/\gamma]{=}[s/r]$ plotted vertically against $r$=(folding number).
See text for further details}
\label{fig:phi21plot}
\end{figure}

We show a selection of our results (up to folding number 50 and
particle number 5) in figure \ref{fig:phi21plot}.

The small (red) points are values which we find definitely to be `non-real'.
The remaining large (black) points are values which appear to have
real spectra up to and including five-particle states.
The solid (red, blue and purple) lines 
are the first three pairs of real series discussed in point 3 below
and the dashed (green) lines are models for which a TBA equation for
the ground state is known (see below).

We find a complicated pattern of results, with increasingly more
models being ruled out by higher and higher particle number transfer
matrices. Since we have only investigated transfer matrices involving
at most 5 particles, our results are somewhat sketchy but they show
the following patterns:

\begin{enumerate}

\item{}
All theories for which the folded models have real spectra are
   obviously real after RSOS restriction, that is those with 
$|\qp|=(1 \pm 1/n)/2$. For the range of $\gamma$ allowed, this gives
only the models 
  $\cM_{4m,2m+1}+\phi_{12}\equiv \cM_{2m+1,4m} + \phi_{21}$
and the models  
  $\cM_{2m+1,m+1}+\phi_{12}\equiv \cM_{m+1,2m+1} + \phi_{21}$.

\item{}
The perturbations of the unitary minimal models have manifestly
unitary $S$--matrices and real spectra, corresponding to the models
$\cM_{p+1,p}+\phi_{12} \equiv \cM_{p,p+1}+\phi_{21}$.

\item{}
It appears that several other infinite series of models may also have
real spectra, at least this appears to be the case from our numerical
tests of the transfer matrices up to 5 particles.
These models form sequences which tend to the `real' theories
described in 1.\ above.
In fact the first example of such a sequence are the unitary models
which tend towards the value $\pi/\gamma=1$, which is the first `real'
model with $n=1$.

The next examples are the series
$\cM_{4k\pm 1,3k\pm 1  }+\phi_{12}$ tending towards $\qp=3/4$,
$\cM_{3k\pm 1,2k\pm 1  }+\phi_{12}$ tending towards $\qp=2/3$,
$\cM_{8k\pm 3,5k\pm 2  }+\phi_{12}$ tending towards $\qp=5/8$,
and so on...

These three pairs of series and the unitary models are shown on figure
\ref{fig:phi21plot} as solid lines.

\end{enumerate}

Some of these models have been considered before in various contexts.
The first of these theories, 
  $\cM_{3,5} +\phi_{21}$, 
was considered by G.~Mussardo in \cite{m35},
where he noted that its spectrum was real despite the fact that the
$S$-matrix was `non-unitary'. 

Since then, (at least) three series of models have been conjectured to
have a ground state described by real TBA equations%
\footnote{We thank R.~Tateo for pointing this out to us.}%
, these being 
$\cM_{m+1,2m+1} + \phi_{21}$ \cite{RST94,Martins92},
$\cM_{2m+1,4m} + \phi_{21}$ \cite{Melzer94},
and $\cM_{2m+1,4m-2} + \phi_{21}$ \cite{DDT00}.
These are massive counterparts of the the $|I|$=1, 2 and 4 massless
flows considered in \cite{DDT00} with $\pi/\gamma = 1/2 + |I|/(2r)$
and shown in figure \ref{fig:phi21plot} as dashed (green) lines.
However, as is well known, simply knowing the TBA equations for the
ground state does not determine the excited state spectrum uniquely,
(a particular example of this effect being the `type II' ambiguity
noted in \cite{12_15}) and the scaling function can describe the
ground state of a genuine unitary model and also a `non-real'
non-unitary model. So, while TBA equations are a useful alternative
way to study the spectrum of a theory, it is not easy to deduce
reality properties from the TBA equation for a ground state.

\subsubsection{The $\phi_{15}$ perturbations with $\pi/\gamma > 1$}

These theories again have at least one higher kink $K_1$ and again we
find that the most stringent restrictions already arise from the
two-particle transfer matrix involving $K_0$ and $K_1$.

Our numerical analysis shows that only two series of models have the
possibility to be \real.  These are the theories with $p'=2p\pm
1\,\bmod\,4p$ for which the $S$--matrix is Hermitian analytic which
guarantees the reality of the spectrum at the level of the Bethe-Yang
equations, and the theories with $p'=\pm 1\,\bmod\,4p$ for which the
reality of the spectrum is mysterious.

\subsubsection{The $\phi_{15}$ perturbations with $\pi/\gamma < 1$}

Again, these theories have no higher kinks in their spectra, and at
most one breather, and the $S$--matrix bootstrap closes on these
particles. The only possible violation of reality of the spectrum
could be introduced by the fundamental kink $S$--matrix, since the 
kink-breather and breather-breather $S$-matrices are pure phases.

\begin{figure}[h]
{\par\centering \includegraphics{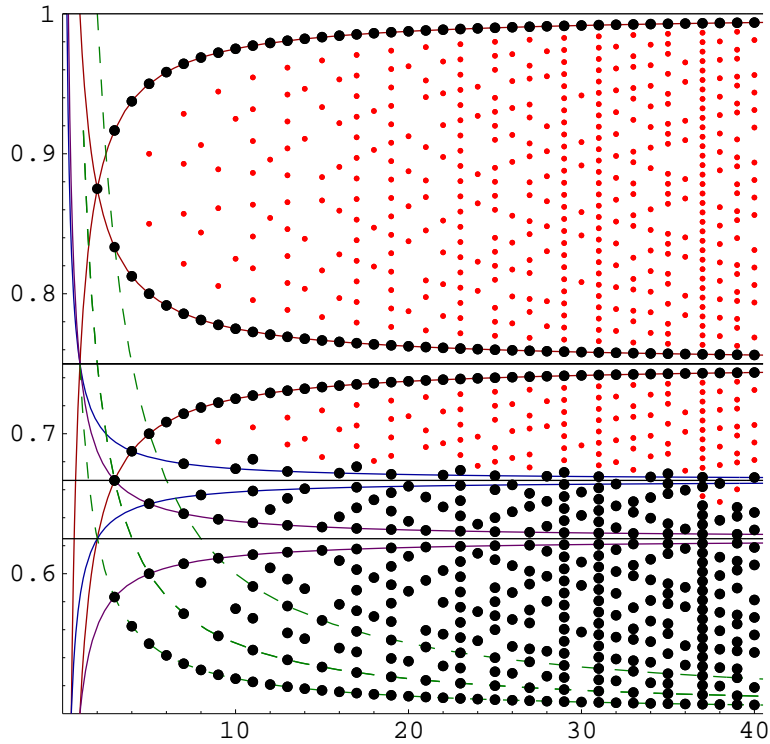} \par}

\caption{\small Numerical results for the eigenvalues of the transfer
matrices of $\cM_{rs} + \phi_{15}$.
The heavy (black) points are models for which the multi-particle
spectrum appears to be real up to 4 particles, and the light (red)
points those which are non-real, with 
$\pi/\gamma{=}s/4r$ plotted vertically against $r$=(folding number)/2.
See text for further details}
\label{fig:phi15plot}
\end{figure}

We find an even more complicated pattern of results than for the
$\phi_{12}$ perturbations, once again with increasingly more
models being ruled out by higher and higher particle number transfer
matrices. In this case the size of the transfer matrices is much
larger, and we have only investigated transfer matrices involving
at most 4 particles. Our results are even more sketchy but they show
the following patterns:

\begin{enumerate}

\item{}
All theories for which the folded models have real spectra are
  still obviously real after RSOS restriction, that is those with 
$|\qp|=(1 \pm 1/n)/2$. For the range of $\gamma$ allowed, this gives
only the models 
  $\cM_{p,2p+1}+\phi_{15}$
and
  $\cM_{2p+1,4p+4}+\phi_{15}$.

\item{}
Again, it appears that several other infinite series of models may
also have 
real spectra, at least this appears to be the case from our numerical
tests of the transfer matrices up to 4 particles.
These models form sequences which tend to the `real' theories
described in 1.\ above.

The first example are the theories,
$\cM_{k,4k - 1  }+\phi_{15}$ tending towards $\qp=1$ 
and the series
$\cM_{k,3k\pm 1  }+\phi_{15}$ tending towards $\qp=3/4$.

At the next level, we find that up to four particles, there are four
infinite series tending to $\qp=2/3$, namely $\cM_{3k\pm 1,8k\pm 3
}+\phi_{15}$ and $\cM_{6k\pm 1,16k\pm 2 }+\phi_{15}$.  Since the
existence of this second series is in some sense a new phenomenon, we
checked that it survives our numerical tests at the five particle
transfer matrix level, but we have no opinion whether or not it will
survive at all higher particle numbers.

\end{enumerate}

Again, three infinite series of these models have been considered
before in various contexts \cite{DDT00}.
These series are exactly those which are related to the infinite
series of $\phi_{21}$ perturbations by interchanging the two terms in
the ZMS potential as described in e.g. \cite{12_15}:
\begin{eqnarray}
  \cM_{2m+1,m+1} + \phi_{12}
&\leftrightarrow&
  \cM_{2m+1,4m+4} + \phi_{15}
\;,\nonumber\\
  \cM_{4m,2m+1} + \phi_{12}
&\leftrightarrow&
  \cM_{m,2m+1} + \phi_{15}
\;,\\
  \cM_{4m-2,2m+1} + \phi_{12}
&\leftrightarrow&
  \cM_{2m-1,4m+2} + \phi_{15}
\;.
\nonumber
\end{eqnarray}
These series are shown as dashed (green) lines in figure
\ref{fig:phi15plot}.
The final pair of theories is particularly interesting as being an
example of a `type II' pair in that these are different theories but
share exactly the same ground state scaling function and have a
common sub-sector of multi-particle states with identical finite size
energy corrections.

\section{Conclusions}
\label{sec:conclusions}

We have investigated the finite volume spectra of theories based on
\aoo and \att imaginary coupled affine Toda field theories. Our main
results can be summarised as follows:

\begin{enumerate}
\item{} All theories (original, folded, RSOS) based on \aoo\ have real
spectra. To show this, we used the fact that the finite volume
spectrum of the RSOS theories can be obtained as suitable projections
of folded theories, which are in turn manifestly unitary QFTs related
to sine-Gordon theory. We also presented new evidence for this
(previously known) correspondence between the folded and RSOS models
based on the transfer matrix method.
\item{} We conjectured that a similar correspondence exists between
folded \att\ models and RSOS models based on \att, and supported this
by numerical diagonalisation of their transfer matrices.
\item{} We presented substantial evidence that unrestricted (both
the folded and the original) \att\ models have complex finite volume
spectra in general, perhaps with the exception of a few special values
of the coupling constant.
\item{} Similarly, it seems that RSOS theories based on \att\ have
complex spectra in general, with the exception of some special
sequences asymptotically approaching the special values of the
coupling for which the unrestricted models may have real spectrum.
These sequences in particular include the perturbations of the unitary
minimal models, for which we know that the spectrum is real.
\end{enumerate}

There are quite a few open questions remaining. First of all, the NLIE
description must be extended to excited states of \att\ related
models; this could provide us conclusive evidence on whether or not
the spectra of the unrestricted \att\ theories are real for the
special values of the coupling constants for which the transfer matrix
calculations found no violation of reality. On the other hand, our
results that show that the spectrum is complex in general must be
reproduced by such an extension of the NLIE.

Second, the projection of the folded spectrum to the RSOS spectrum
must be explicitly realized. Together with the extension of the NLIE
to excited states, this would open the way to (a) a systematic
description of the spectra of the RSOS models; (b) determining whether
the sequences for which we found no violation of reality with our
methods really have real spectra.

It seems appropriate to mention that another interesting problem is
the relation of the \att\ model and `$\phi_{21}$' perturbations to the
corresponding $q$-state Potts models. See \cite{DPT02} for 
TBA equations for the ground states of these models and for a 
discussion of the particle spectra in the $q$-state Potts models.

Third, as we mentioned in the text, the full spectrum of \att\ related
theories is not yet closed in full generality. An example is
$\cM_{3,5}+\phi_{12}$, for which the closed spectrum is not known up
to date. One could attempt to close the $S$ matrix bootstrap (we know
of some attempts that failed, and we ourselves tried unsuccessfully);
but equally well, it could possibly be found by extending the NLIE to
describe the excited state spectrum of \att\ related theories.

{\bf Acknowledgments}

The authors would like to thank P.~Dorey, H.~Saleur and R.~Tateo for
valuable comments and discussions. We also thank G.~Mussardo for
providing a correct set of 
$q$-$6j$ symbols for the $\phi_{12}$ perturbations. This work was
supported by a Royal Society joint project grant. G.T. is supported
by a Zolt\'an Magyary Fellowship from the Hungarian Ministry of
Education, and is partly supported from Hungarian funds OTKA
T029802/99 and FKFP-0043/2001. G.T. is also grateful to the Department
of Mathematics, King's College, London for their hospitality. 

\section*{Appendix}
\appendix
\makeatletter 
\renewcommand\theequation{\hbox{\normalsize\Alph{section}.\arabic{equation}}}
\makeatother

\section{\label{rsos12} Scattering amplitudes for \protect$ \mathcal{M}_{p,p'}+\Phi_{1,2}\protect $}

The scattering amplitudes for the fundamental kinks in $ \mathcal{M}_{p,p'}+\Phi_{1,2} $
were written down by Smirnov in \cite{smirnov}. Here we briefly recall his
result and give explicit expressions for the necessary $ q $-$ 6j $ symbols
as it seems there are none available in the literature that are free
of misprints 
and errors.

\begin{figure}[h]
\psfrag{0}{$0$}
\psfrag{1}{$1$}
\psfrag{2}{$2$}
\psfrag{b}{$j_{\mathrm max}-1$}
\psfrag{l}{$j_{\mathrm max}$}
\psfrag{oh}{$\frac{1}{2}$}
\psfrag{th}{$\frac{3}{2}$}
\psfrag{fh}{$\frac{5}{2}$}
\psfrag{bh}{$j_{\mathrm max}-\frac{3}{2}$}
\psfrag{lh}{$j_{\mathrm max}-\frac{1}{2}$}
{\par\centering \includegraphics{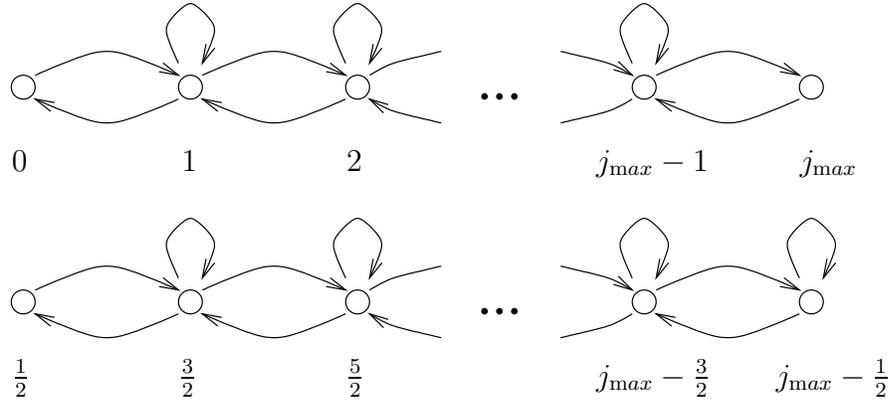} \par}

\caption{Adjacency rules for kinks in \protect$ \Phi_{12}\protect
$ perturbations for $j_{\mathrm max}\in{\mathbb Z}$}
\label{adj12fig}
\end{figure}

The allowed vacuum sequences $ \left\{ j_{1},\ldots ,\, j_{n}\right\}  $
have the following adjacency rules:
\begin{eqnarray}
j_{k+1} & = & 1\; \mathrm{if}\; j_{k}=0\nonumber\\
j_{k+1} & \in  & \left\{ \max \left( 0,\, j_{k}-1\right) \dots \min \left( j_{k}+1,\, p-3-j_{k}\right) \right\} \; \mathrm{if}\; j_{k}\neq 0
\end{eqnarray}
and are of two types: either all $ j_{k} $ are integers or all $
j_{k}\in {\mathbb Z}+1/2 $. The maximum value for $ j_{k} $ is $
j_{\mathrm{max}}=(p-2)/2 $. The adjacency rules for these two
sequences are shown on Figure \ref{adj12fig} for the case when $
j_{\mathrm{max}} $ is integer; for the case $ j_{\mathrm{max}}\in
{\mathbb Z}+1/2$ two very similar graphs result. Let us introduce the
notation
\begin{equation}
\label{xi12}
q=\exp \left( \frac{i\pi p'}{p}\right) \, ,\quad x=\exp \left( \frac{2\pi \vartheta }{\xi }\right) \quad \mathrm{where}\quad \xi =\frac{2}{3}\frac{\pi p}{2p'-p}\,,
\end{equation}
and the $q$-numbers
\begin{equation}
[n]=\frac{q^n-q^{-n}}{q-q^{-1}\,.}
\end{equation}
Then the 2-kink scattering amplitudes take the form
\begin{eqnarray}
S^{cd}_{ab}(\vartheta ) & = & \hat{S}^{cd}_{ab}\left( x,q\right) \, S_{0}(\vartheta )\nonumber \\
\hat{S}^{cd}_{ab}\left( x,q\right)  & = & \frac{\left( q^{2}-1\right) \left( q^{3}+1\right) }{q^{\frac{5}{2}}}\delta_{bd}+\left( \frac{1}{x}-1\right) q^{\frac{3}{2}-\frac{1}{2}\left( C_{b}+C_{d}-C_{a}-C_{c}\right) }\left\{ \begin{array}{ccc}
1 & c & d\\
1 & a & b
\end{array}\right\}_{q}+\nonumber \\
 &  & \left( 1-x\right) q^{-\frac{3}{2}+\frac{1}{2}\left( C_{b}+C_{d}-C_{a}-C_{c}\right) }\left\{ \begin{array}{ccc}
1 & c & d\\
1 & a & b
\end{array}\right\}_{q}\label{2kink_12} 
\end{eqnarray}
where 
\begin{eqnarray}
S_{0}(\vartheta ) & = & \pm \frac{1}{4i}\left( \sinh \frac{\pi }{\xi }\left( \vartheta -\pi i\right) \sinh \frac{\pi }{\xi }\left( \vartheta -\frac{2\pi i}{3}\right) \right) ^{-1}\times \nonumber \\
 &  & \exp \left( -2i\int_{0}^{\infty }\frac{\sin k\vartheta \sinh \frac{\pi k}{3}\cosh \left( \frac{\pi }{6}-\frac{\xi }{2}\right) k}{k\cosh \frac{\pi k}{2}\sinh \frac{\xi k}{2}}dk\right) \: .\label{zms_scalar_factor} 
\end{eqnarray}
 
\begin{equation}
C_{a}=a(a+1)\end{equation}
and the $ q $-$ 6j $ symbols take the form
\begin{eqnarray}
 &  & \left\{ \begin{array}{ccc}
1 & a-2 & a-1\\
1 & a & a-1
\end{array}\right\}_{q}=\left\{ \begin{array}{ccc}
1 & a+2 & a+1\\
1 & a & a+1
\end{array}\right\}_{q}=1\nonumber\\
 &  & \left\{ \begin{array}{ccc}
1 & a+1 & a+1\\
1 & a & a
\end{array}\right\}_{q}=\left\{ \begin{array}{ccc}
1 & a & a\\
1 & a+1 & a+1
\end{array}\right\}_{q}=\frac{\sqrt{[2a+4][2a]}}{[2a+2]}\nonumber\\
 &  & \left\{ \begin{array}{ccc}
1 & a & a+1\\
1 & a & a+1
\end{array}\right\}_{q}=\frac{[2]}{[2a+1][2a+2]}\nonumber\\
 &  & \left\{ \begin{array}{ccc}
1 & a & a-1\\
1 & a & a-1
\end{array}\right\}_{q}=\frac{[2]}{[2a][2a+1]}\nonumber\\
 &  & \left\{ \begin{array}{ccc}
1 & a+1 & a\\
1 & a & a+1
\end{array}\right\}_{q}=\left\{ \begin{array}{ccc}
1 & a & a+1\\
1 & a+1 & a
\end{array}\right\}_{q}=\frac{\sqrt{[2a+4][2a]}}{[2a+2]}\nonumber\\
 &  & \left\{ \begin{array}{ccc}
1 & a & a-1\\
1 & a & a+1
\end{array}\right\}_{q}=\left\{ \begin{array}{ccc}
1 & a & a+1\\
1 & a & a-1
\end{array}\right\}_{q}=\frac{\sqrt{[2a-1][2a+3]}}{[2a+1]}\nonumber\\
 &  & \left\{ \begin{array}{ccc}
1 & a & a\\
1 & a & a+1
\end{array}\right\}_{q}=\left\{ \begin{array}{ccc}
1 & a & a+1\\
1 & a & a
\end{array}\right\}_{q}=-\frac{[2]}{[2a+2]}\sqrt{\frac{[2a+3]}{[2a+1]}}\nonumber\\
 &  & \left\{ \begin{array}{ccc}
1 & a & a\\
1 & a+1 & a
\end{array}\right\}_{q}=\left\{ \begin{array}{ccc}
1 & a+1 & a\\
1 & a & a
\end{array}\right\}_{q}=\frac{[2]}{[2a+2]}\nonumber\\
 &  & \left\{ \begin{array}{ccc}
1 & a & a\\
1 & a & a-1
\end{array}\right\}_{q}=\left\{ \begin{array}{ccc}
1 & a & a-1\\
1 & a & a
\end{array}\right\}_{q}=\frac{[2]}{[2a]}\sqrt{\frac{[2a-1]}{[2a+1]}}\nonumber\\
 &  & \left\{ \begin{array}{ccc}
1 & a & a\\
1 & a-1 & a
\end{array}\right\}_{q}=\left\{ \begin{array}{ccc}
1 & a-1 & a\\
1 & a & a
\end{array}\right\}_{q}=-\frac{[2]}{[2a]}\nonumber\\
 &  & \left\{ \begin{array}{ccc}
1 & a & a\\
1 & a & a
\end{array}\right\}_{q}=\frac{[2a-1][2a+3]-1}{[2a][2a+2]}
\label{rsos12_6j}\end{eqnarray}

We remark that the square roots in (\ref{rsos12_6j}) carry a sign ambiguity
which must be resolved by an appropriate choice of the branches of the square
root function. One must consider the square roots of each $ q $-numbers separately
and fix a sign for the expressions

\begin{equation}
\sqrt{[n]}\, ,\quad n=1,2,\dots \end{equation}
 and then use the selected representative consistently in all formulas. Such
a choice of branch is necessary in order for the amplitudes to satisfy the Yang-Baxter
equation, $ R $-matrix unitarity and appropriate crossing relations.

\section{\label{rsos15} Scattering amplitudes for \protect$ \mathcal{M}_{p,p'}+\Phi_{1,5}\protect $}

In this appendix we present a corrected version of the S-matrix of 
$\Phi_{1,5} $ perturbations written down in the paper \cite{rsos15} by
one of the authors.  The original formulas have misprints and some of
them have the wrong normalisation factors. The correct ones can be
obtained by imposing crossing symmetry and RU on the amplitudes. The
choice of the normalisation factors amount to defining the scalar
product of the multi-kink states correctly (i.e. satisfying the
constraints imposed by the quantum group symmetry).

\begin{figure}
\psfrag{0}{$0$}
\psfrag{1}{$\frac{1}{2}$}
\psfrag{2}{$1$}
\psfrag{b}{$j_{\mathrm max}-\frac{1}{2}$}
\psfrag{l}{$j_{\mathrm max}$}
{\par\centering \includegraphics{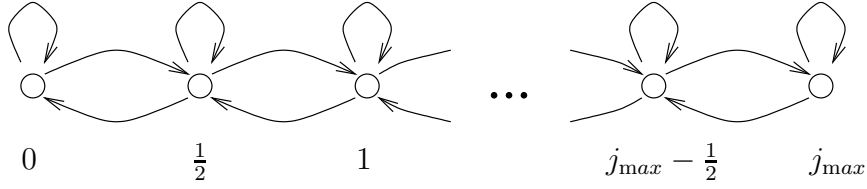} \par}

\caption{Adjacency rules for kinks in \protect$ \Phi_{15}\protect $ perturbations}
\label{adj15fig}
\end{figure}

In $\Phi_{1,5} $ perturbations, the allowed vacuum sequences are
composed of highest weights of the group $ U_{q^{4}}(sl(2)) $ where
\begin{equation}
q=\exp \left( \frac{i\pi p'}{4p}\right) 
\, .\end{equation} 
They are labelled by $ j_{k}=0,1/2,\ldots ,\, j_{max} $ with $
j_{max}=(p-2)/2 $.  The adjacency rules for a sequence $
\{j_{1},\ldots ,\, j_{n}\} $ are $ \left| j_{k+1}-j_{k}\right| =0 $ or
$ 1/2 $ (Figure \ref{adj15fig}). The zero difference corresponds to a
neutral kink, while the nonzero to charged ones.
Let us introduce the notation 
\begin{eqnarray}
y & = & \exp \left( \frac{\pi }{\xi }\vartheta \right) \, ,\quad \xi =\frac{4}{3}\frac{\pi p}{p'-2p}\, ,\nonumber\\
\left[ z\right]_{4} & = & \frac{q^{4z}-q^{-4z}}{q^{4}-q^{-4}}\, .
\end{eqnarray}
 The scattering amplitudes for charged kinks are (with a sign corrected with
respect to \cite{rsos15}) 
\begin{eqnarray}
S^{cd}_{ab} & = & \left( -\left( \frac{y^{2}}{q}-\frac{q}{y^{2}}-\frac{1}{q}+q\right) \delta_{ac}\left( \frac{[2b+1]_{4}[2d+1]_{4}}{[2a+1]_{4}[2c+1]_{4}}\right) ^{1/2}\right. \nonumber\\
 & + & \left. \left( \frac{y^{2}}{q^{5}}-\frac{q^{5}}{y^{2}}-\frac{1}{q}+q\right) \delta_{bd}\right) S_{0}(\vartheta )\, \, ,
\end{eqnarray}
 where $ S_{0}(\vartheta ) $ is the function defined in (\ref{zms_scalar_factor}).
For the sign ambiguities occurring as a result of square roots of $ q $-numbers
see the remark at the end of Appendix \ref{rsos12}.

The remaining amplitudes include neutral kinks. The neutral
kink-neutral kink scattering, neutral kink-charged kink forward
scattering and neutral kink-charged kink reflection were correct in
the original paper \cite{rsos15} and are the following

\begin{eqnarray}
S^{aa}_{aa} & = & \frac{q^{6}y^{2}+y^{2}q^{8}-q^{8}-q^{4}y^{2}+y^{2}-q^{10}y^{2}+y^{4}q^{2}-y^{2}q^{2}}{y^{2}q^{5}}S_{0}(\vartheta )\, ,\nonumber\\
S^{ba}_{ab} & = & \frac{(y^{2}+q^{6})(y^{2}-1)}{y^{2}q^{3}}S_{0}(\vartheta )\, ,\nonumber\\
S^{ba}_{aa} & = & -\frac{(q^{4}-1)(y^{2}+q^{6})}{yq^{5}}S_{0}(\vartheta )\, .
\end{eqnarray}
The amplitudes describing two charged kinks turning into two neutral
ones were incorrectly normalised (they include a nontrivial
Clebsh-Gordan in the unrestricted theory \cite{rsos15}). The correct
form is
\begin{equation}
\displaystyle S^{aa}_{ab}=i\frac{c_{aa}}{c_{ba}}\frac{(q^{4}-1)(y^{2}-1)}{q^{2}y}
\end{equation}
 For the reverse process the amplitude is 
\begin{equation}
\displaystyle S^{ab}_{aa}=i\frac{c_{ab}}{c_{aa}}\frac{(q^{4}-1)(y^{2}-1)}{q^{2}y}
\end{equation}
 where the normalisation factors $ c_{ab} $ are 
\begin{equation}
c_{ab}=\left\{ \begin{array}{ll}
\alpha_{1}(-1)^{2a}\sqrt{\frac{[2b+1]_{4}}{[2a+1]_{4}}} & a=b\pm \frac{1}{2}\\
\alpha_{2} & a=b
\end{array}\right. \, .\end{equation}
$ \alpha_{1,2} $ are constants which are left free by the constraints
of RU and crossing. The $ c_{ab} $ appear in the crossing relations,
which take the form
\begin{equation}
S^{cd}_{ab}(\vartheta )=\frac{c_{ad}}{c_{bc}}S^{bc}_{da}(i\pi -\vartheta )\, .\end{equation}

\section{\label{transfermat}Transfer matrices and the Bethe-Yang equations}

\subsection{\label{rsos_tm}RSOS transfer matrices}

When we put the theory on a cylindrical space-time with spatial volume
$ L $, the allowed sequences of vacua (see appendices \ref{rsos12} and
\ref{rsos15} for $\phi_{12}$ and $\phi_{15}$ perturbations,
respectively, and \cite{reshetikhin_smirnov} for the $\phi_{13}$
case) are further restricted by the condition $ a_{1}=a_{N+1} $, where
$N$ is the number of particles. We take all particles to be point-like
and ignore all vacuum polarisation contributions. Let us define the
following (generalised) transfer matrix

\begin{center}

\setlength{\unitlength}{3947sp}%

\begingroup\makeatletter\ifx\SetFigFont\undefined%

\gdef\SetFigFont#1#2#3#4#5{%

  \reset@font\fontsize{#1}{#2pt}%

  \fontfamily{#3}\fontseries{#4}\fontshape{#5}%

  \selectfont}%

\fi\endgroup%

\begin{picture}(3825,1062)(376,-511)

\thinlines

\put(2101,539){\line( 0,-1){750}}

\put(3751,539){\line( 0,-1){750}}

\put(1801,164){\line( 1, 0){900}}

\put(2551,539){\line( 0,-1){750}}

\put(4051,164){\line(-1, 0){900}}

\put(3301,539){\line( 0,-1){750}}

\put(-700, 89){\makebox(0,0)[lb]{\smash{\SetFigFont{12}{14.4}{\familydefault}{\mddefault}{\updefault}$T_{a_1a_2\dots a_N}^{b_1b_2\dots b_N}\left(\vartheta |\vartheta_1,\vartheta_2,\dots ,\vartheta_N\right)=$}}}

\put(1801,239){\makebox(0,0)[lb]{\smash{\SetFigFont{12}{14.4}{\familydefault}{\mddefault}{\updefault}$b_1$}}}

\put(2251,239){\makebox(0,0)[lb]{\smash{\SetFigFont{12}{14.4}{\familydefault}{\mddefault}{\updefault}$b_2$}}}

\put(3901,239){\makebox(0,0)[lb]{\smash{\SetFigFont{12}{14.4}{\familydefault}{\mddefault}{\updefault}$b_1$}}}

\put(1801,-136){\makebox(0,0)[lb]{\smash{\SetFigFont{12}{14.4}{\familydefault}{\mddefault}{\updefault}$a_1$}}}

\put(2851, 89){\makebox(0,0)[lb]{\smash{\SetFigFont{20}{24.0}{\familydefault}{\mddefault}{\updefault}...}}}

\put(2251,-136){\makebox(0,0)[lb]{\smash{\SetFigFont{12}{14.4}{\familydefault}{\mddefault}{\updefault}$a_2$}}}

\put(3451,-136){\makebox(0,0)[lb]{\smash{\SetFigFont{12}{14.4}{\familydefault}{\mddefault}{\updefault}$a_N$}}}

\put(3901,-136){\makebox(0,0)[lb]{\smash{\SetFigFont{12}{14.4}{\familydefault}{\mddefault}{\updefault}$a_1$}}}

\put(3451,239){\makebox(0,0)[lb]{\smash{\SetFigFont{12}{14.4}{\familydefault}{\mddefault}{\updefault}$b_N$}}}

\put(2026,-511){\makebox(0,0)[lb]{\smash{\SetFigFont{12}{14.4}{\familydefault}{\mddefault}{\updefault}$\vartheta_1$}}}

\put(2476,-511){\makebox(0,0)[lb]{\smash{\SetFigFont{12}{14.4}{\familydefault}{\mddefault}{\updefault}$\vartheta_2$}}}

\put(3676,-511){\makebox(0,0)[lb]{\smash{\SetFigFont{12}{14.4}{\familydefault}{\mddefault}{\updefault}$\vartheta_N$}}}

\put(4151, 89){\makebox(0,0)[lb]{\smash{\SetFigFont{12}{14.4}{\familydefault}{\mddefault}{\updefault}$\vartheta$}}}

\end{picture}

\end{center} which translates to 
\begin{equation}
T\left( \vartheta |\vartheta_{1},\vartheta_{2},\ldots ,\vartheta
_{N}\right) ^{b_{1}b_{2}\ldots b_{N}}_{a_{1}a_{2}\ldots a_{N}}=\prod
^{N}_{j=1}S^{a_{j+1}b_{j+1}}_{b_{j}a_{j}}\left( \vartheta -\vartheta
_{j}\right) 
\label{RSOS_tm}
\end{equation} 
with the identification $ a_{N+1}\equiv a_{1}\, ,\, b_{N+1}\equiv
b_{1} $.  In the following we shall not always write down the matrix
indices explicitly. From the Yang-Baxter equation, it is straightforward to
prove that these transfer matrices form a commuting family

\begin{equation}
\label{commuting_transfer_matrices}
\left[ T\left( \vartheta |\vartheta_{1},\ldots ,\vartheta_{N}\right) ,T\left( \vartheta '|\vartheta_{1},\ldots ,\vartheta_{N}\right) \right] =0
\end{equation}
We define the following specialised transfer matrices
\begin{eqnarray}
T_{k}\left( \vartheta_{1},\vartheta_{2},\ldots ,\vartheta
_{N}\right) ^{b_{1}b_{2}\ldots b_{N}}_{a_{1}a_{2}\ldots a_{N}} & = &
 T\left( \vartheta_{k}|\vartheta_{1},\vartheta_{2},\ldots
 ,\vartheta_{N}\right) ^{b_{1}b_{2}\ldots b_{N}}_{a_{1}a_{2}\ldots
 a_{N}}\nonumber \\
 & = & (-1)^{\delta }\delta ^{b_{k+1}}_{a_{k}}\prod_{j\neq k}S^{a_{j+1}b_{j+1}}_{b_{j}a_{j}}\left( \vartheta_k -\vartheta_{j}\right) 
\end{eqnarray}
Apart from a phase coming from a plane wave factor of rapidity $ \vartheta_{k} $
in the wave function, this gives the monodromy corresponding to taking the $ k $th
kink around the spatial circle multiplied by a factor $ (-1)^{\delta } $.
The total phase must equal $ (-1)^{F} $ depending on the statistics of the
particle $ k $ ($ F=1 $ for fermions $ F=0 $ for bosons). Thus we obtain
the so-called Bethe-Yang equations \cite{klassen_melzer1}:
\begin{equation}
\label{bethe-yang}
\exp \left( im_{k}R\sinh \vartheta_{k}\right) T_{k}\left( \vartheta_{1},\vartheta_{2},\ldots ,\vartheta_{N}\right) ^{b_{1}b_{2}\ldots b_{N}}_{a_{1}a_{2}\ldots a_{N}}\Psi ^{a_{1}a_{2}\ldots a_{N}}=(-1)^{F+\delta }\Psi ^{b_{1}b_{2}\ldots b_{N}}
\end{equation}
where $ \Psi ^{a_{1}a_{2}\ldots a_{N}} $ is the wave function amplitude,
defined by the decomposition of the state $ \left| \Psi \right\rangle  $
as follows
\begin{equation}
\left| \Psi \right\rangle =\sum_{a_{1},\ldots ,a_{N}}\Psi ^{a_{1}a_{2}\ldots a_{N}}\left| K_{a_{1}a_{2}}\left( \vartheta_{1}\right) \ldots K_{a_{N}a_{1}}\left( \vartheta_{N}\right) \right\rangle \, .\end{equation}
The energy and the momentum of the state (relative to the vacuum) are given by 
\begin{equation}
 E = \sum ^{N}_{k=1}m_{k}\cosh \vartheta_{k}\, ,\, 
 P=\sum^{N}_{k=1}m_{k}\sinh \vartheta_{k}\, \, .
\end{equation} 
Due to the commutation relation (\ref{commuting_transfer_matrices}),
the equations (\ref{bethe-yang}) for the vector $ \Psi $ are
compatible and can be reduced to scalar equations by simultaneously
diagonalising the transfer matrices. Let us denote the eigenvalues of
$ T\left( \vartheta |\vartheta_{1},\ldots ,\vartheta_{N}\right) $ by
$ \lambda ^{(s)}\left( \vartheta |\vartheta_{1},\ldots ,\vartheta
_{N}\right) $ with the corresponding eigenvectors $ \psi ^{(s)}\left(
\vartheta_{1},\ldots ,\vartheta_{N}\right) $ ($ s $ is just an index
enumerating the eigenvalues and the eigenvectors can be chosen
independent of $ \vartheta $ due to the commutativity
(\ref{commuting_transfer_matrices})).  Then the solutions of the
Bethe-Yang equations (\ref{bethe-yang}) are given by
\begin{equation}
\Psi ^{a_{1}a_{2}\ldots a_{N}}=\psi ^{(s)}\left( \vartheta_{1},\ldots ,\vartheta_{N}\right) ^{a_{1}a_{2}\ldots a_{N}}\end{equation}
where the rapidities solve the scalar Bethe Ansatz equations
\begin{equation}
\label{scalar_bethe_eqns}
\exp \left( im_{k}R\sinh \vartheta_{k}\right) \lambda ^{(s)}\left( \vartheta_{k}|\vartheta_{1},\ldots ,\vartheta_{N}\right) =(-1)^{F+\delta }
\end{equation}

\subsection{\label{folded_tm} Folded transfer matrices}

\begin{figure}[hbt]
\begin{center}
\setlength{\unitlength}{3000sp}%
\begingroup\makeatletter\ifx\SetFigFont\undefined%
\gdef\SetFigFont#1#2#3#4#5{%
  \reset@font\fontsize{#1}{#2pt}%
  \fontfamily{#3}\fontseries{#4}\fontshape{#5}%
  \selectfont}%
\fi\endgroup%
\begin{picture}(1440,1755)(451,-1141)
\thinlines
{\put(721,299){\line( 1,-1){1080}}
}%
{\put(721,-781){\line( 1, 1){1080}}
}%
\put(721,-331){\makebox(0,0)[lb]{\smash{\SetFigFont{12}{14.4}{\rmdefault}{\mddefault}{\itdefault}{a}%
}}}
\put(1216,-781){\makebox(0,0)[lb]{\smash{\SetFigFont{12}{14.4}{\rmdefault}{\mddefault}{\itdefault}{b}%
}}}
\put(1666,-331){\makebox(0,0)[lb]{\smash{\SetFigFont{12}{14.4}{\rmdefault}{\mddefault}{\itdefault}{c}%
}}}
\put(1171,119){\makebox(0,0)[lb]{\smash{\SetFigFont{12}{14.4}{\rmdefault}{\mddefault}{\itdefault}{d}%
}}}
\put(1891,479){\makebox(0,0)[lb]{\smash{\SetFigFont{12}{14.4}{\rmdefault}{\mddefault}{\itdefault}{$Q_{dc}$}%
}}}
\put(1891,-1141){\makebox(0,0)[lb]{\smash{\SetFigFont{12}{14.4}{\rmdefault}{\mddefault}{\itdefault}{$Q_{bc}$}%
}}}
\put(451,-1141){\makebox(0,0)[lb]{\smash{\SetFigFont{12}{14.4}{\rmdefault}{\mddefault}{\itdefault}{$Q_{ab}$}%
}}}
\put(451,479){\makebox(0,0)[lb]{\smash{\SetFigFont{12}{14.4}{\rmdefault}{\mddefault}{\itdefault}{$Q_{ad}$}%
}}}
\end{picture}
\caption{Vacuum labels and topological charges for two-particle
$S$-matrices in folded models}
\label{folded_S}
\end{center}
\end{figure}
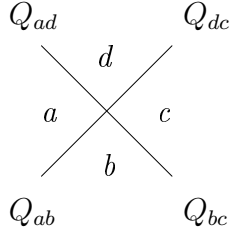

For folded models the vacua are labelled by an integer $a$ modulo the
folding number $k$. The allowed sequences satisfy
\begin{equation}
a_{i+1}=a_{i}+Q\ \bmod\, k
\end{equation}
where $Q$ are the possible topological charges of the solitons 
\begin{equation}
Q=\left\{
\begin{array}{ll}
\pm 1, & \mathrm{sine-Gordon} \\
+1,\, 0,\, -1 & \mathrm{ZMS}
\end{array}
\right.
\end{equation}
In finite volume we require $ a_{1}=a_{N+1} \ \bmod\, k$ for periodic
boundary conditions. Then the transfer matrix has the same form as in
equation (\ref{RSOS_tm}) except that now the matrix $S$ is constructed
from the scattering matrix $\tilde S$ of the sine-Gordon/ZMS model in
the following way:
\begin{equation}
S_{ab}^{cd}(\vartheta )={\tilde S}_{Q_{ab},Q_{bc}}^{Q_{ad},Q_{dc}}(\vartheta )
\end{equation}
where $Q_{ab}$ denotes the charge of the soliton connecting the vacua
$a$ and $b$ (see Figure \ref{folded_S}).

\end{document}